\documentclass[12pt]{iopart}

\usepackage{iopams}
\usepackage{dcolumn}
\usepackage{graphicx}
\usepackage{bm}

\newcommand{\BaFA}{BaFe$_{2}$As$_{2}$}
\newcommand{\BaCoP}{Ba(Fe$_{1-y}$Co$_{y}$)$_{2}$(As$_{1-x}$P$_x$)$_{2} $}
\newcommand{\BadP}{Ba(Fe$_{0.97}$Co$_{0.03}$)$_{2}$(As$_{1-x}$P$_x$)$_{2}$}
\newcommand{\BafP}{Ba(Fe$_{0.95}$Co$_{0.05}$)$_{2}$(As$_{1-x}$P$_x$)$_{2}$}
\newcommand{\BasP}{Ba(Fe$_{0.93}$Co$_{0.07}$)$_{2}$(As$_{1-x}$P$_x$)$_{2}$}
\newcommand{\BaCo}{Ba(Fe$_{1-y}$Co$_{y}$)$_{2}$As$_{2}$}
\newcommand{\BaP}{BaFe$_{2}$(As$_{1-x}$P$_x$)$_{2}$}

\pdfoutput=1

\begin{document}

\title[\BaCoP]{The interplay of electron doping and chemical pressure in Ba(Fe$_{1-y}$Co$_{y}$)$_{2}$(As$_{1-x}$P$_x$)$_{2} $}

\author{Veronika Zinth and Dirk Johrendt}

\address{Department Chemie der Ludwig-Maximilians-Universit\"{a}t M\"{u}nchen,\\ Butenandtstr- 5-13 (Haus D), 81377 M\"{u}nchen, Germany}

\ead{johrendt@lmu.de}

\begin{abstract}
The effects of internal chemical pressure on electron doped iron arsenide superconductors are studied in the series {\BaCoP}. Combinations of both dopants induce superconductivity also in such areas where only one would not suffice, and can likewise move the system into an overdoped state, while no higher critical temperature than 31 K in {\BaP} was found. The phase diagram gives no evidence of holes in {\BaP} as suggested by recent photoemission experiments. Chemical and physical pressure act similarly in {\BaCoP}, but our data reveal that the most important control parameter is the length of the Fe-As bond and not the unit cell volume. This emphasizes that differences between chemical and physical pressure which manifest oneself as the non-linear reduction of the Fe-As distance in {\BaP} are strongly linked to the superconducting properties also in the Co doped compounds.
\end{abstract}

\pacs{
 74.70.Xa, 
 74.25.Dw, 
 74.62.Dh, 
 74.62.Fj, 
 }

\section{Introduction}

The phase diagrams of copper-oxide, heavy-fermion-, and iron-arsenide superconductors imply the occurrence of unconventional superconductivity in the proximity of a magnetic quantum critical point \cite{Sachdev-2010,Steglich-2010,Liu-2008,Giovannetti-2011}. In copper-oxides, solely chemical doping can destabilize the antiferromagnetic ground-state in favor of the superconducting one, while in heavy-fermion and iron-based materials doping as well as pressure can induce superconductivity. Instead of applying external pressure, the volume reduction can also be achieved by substitution of elements with those of smaller sizes, which is often referred to as chemical or internal pressure.

Virtually all possible options of suppressing the spin-density-wave (SDW) state have been realized with the iron arsenide {\BaFA} \cite{Rotter-2008-1} as parent compound. Hole-doping by substitution of K for Ba \cite{Rotter-2008-2}, electron-doping by substitution of Co for Fe \cite{Sefat-2008}, applying physical pressure \cite{Alireza-2009}, and finally by internal pressure through replacing smaller P-atoms for As-atoms \cite{Jiang-2009}. We have proposed a more sophisticated approach to chemical pressure in {\BaP}  by showing that the length of the Fe-As bond is a gauge of the magnetic moment, which is gradually suppressed while doping with phosphorus because the \textit{width} of the $3d$-bands increase \cite{Rotter-2010}. On the other hand we have recently studied (Ba$_{1-x}$K$_x$)(Fe$_{0.93}$Co$_{0.07}$)$_2$As$_2$ and were able to compensate electron- and hole-doping, strongly suggesting that the charge of the (FeAs)$^{\delta-}$ layer \textit{viz.} band \textit{filling} is likewise an important control parameter with respect to suppression of the SDW and emerging of superconductivity \cite{Zinth-2011}.

In this paper we study the interplay between electron doping by cobalt and the effect of internal chemical pressure by phosphorus substitution in the series  \BaCoP. We present a phase diagram and analyze the mutual effects of both modifications with respect to inducing or suppressing of superconductivity. Chemical and physical pressure effects on the critical temperature are discussed in terms of the Fe-As bond length as essential control parameter.

\section{Experimental Methods}
%

Fe$_{1-y}$Co$_y$As$_{1-x}$P$_x$ was prepared by heating stoichiometric mixtures of iron, cobalt, arsenic and phosphorus at 973~K for 10~h in sealed silica ampoules under an atmosphere of purified argon, followed by   homogenizing and sintering two times at 1123~K for 30~h. Stoichiometric amounts of Ba were added and the mixtures were heated to 943~K for 10~h in alumina crucibles. The products were homogenized and sintered at 1273~K and 1373~K, respectively, for another 30~h. The dc-resistivities were measured with pellets which have been cold pressed and sintered at 1273~K for 10~h. Bulk superconductivity was confirmed by ac-susceptibility measurements, and the superconducting transition temperatures were extracted by fitting the steepest descent with a line and using the point of intersection with the normal-state susceptibility as $T_c$ (onset).

Powder diffraction data were measured using a Huber G670 diffractometer with Co-K$_{\alpha1}$ or Cu-K$_{\alpha1}$-radiation. Phase homogeneity and structural parameters were determined by Rietveld refinements with the TOPAS package \cite{TOPAS} using the fundamental parameter approach with an empirical 2$\theta$-dependent intensity correction for Guinier geometry.
The Fe:Co ratios were fixed and the As:P ratios refined. Both Co and P contents were checked by EDX measurements for most samples, revealing variations by no more than 1~\% for the Co content and by no more than 2-4~\%  for the P content.  In  a few samples small percentages of Fe$_2$P were observed, while others contained traces of an unknown foreign phase with main reflection at 2$\theta \approx 28^{\circ}$, which was also reported to occur similar samples \cite{Martin-Thes}.

\section{Results and Discussion}
\begin{figure}[h]
\center{
\includegraphics[width=0.85\textwidth]{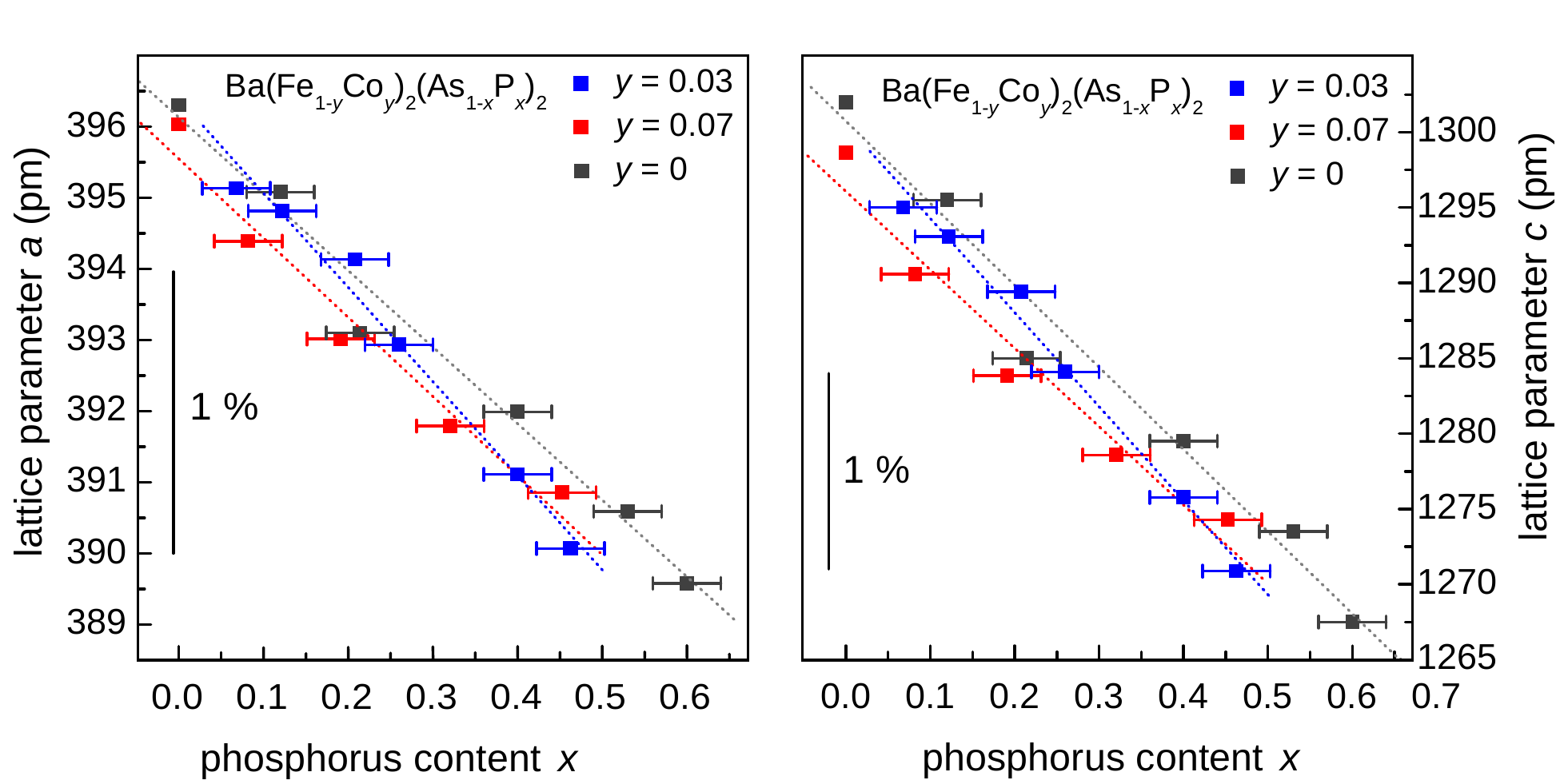}
\caption{{\BaCoP}: lattice parameters $a$ and $c$ (pm) for $y$ = 0 (black) \cite{Martin-Thes}, $y$ = 0.03 (blue) and $y$ = 0.05 (red). }
\label{fig:lattice}
}
\end{figure}

Figure \ref{fig:lattice} shows the variation of the lattice para\-meters $a$ and $c$ with increasing phosphorus substitution for phosphorus-only doped samples (black) \cite{Martin-Thes} and samples with additional 3~\% (blue) or 7~\% cobalt (red). For all Co doping levels both lattice para\-meters decrease with increasing phosphorus substitution, as is expected because of so-called "chemical pressure" by the smaller ionic radius of phosphorus atoms compared to arsenic.

Co doping seems to decrease both lattice para\-meters slightly, but the effect is minimal compared to the decrease due to phosphorus substitution. Thus all structural parameters are dominated by effects of the phosphorus content and hardly influenced by Co doping.

\begin{figure}[h!]
\center{
\includegraphics[width=0.6\textwidth]{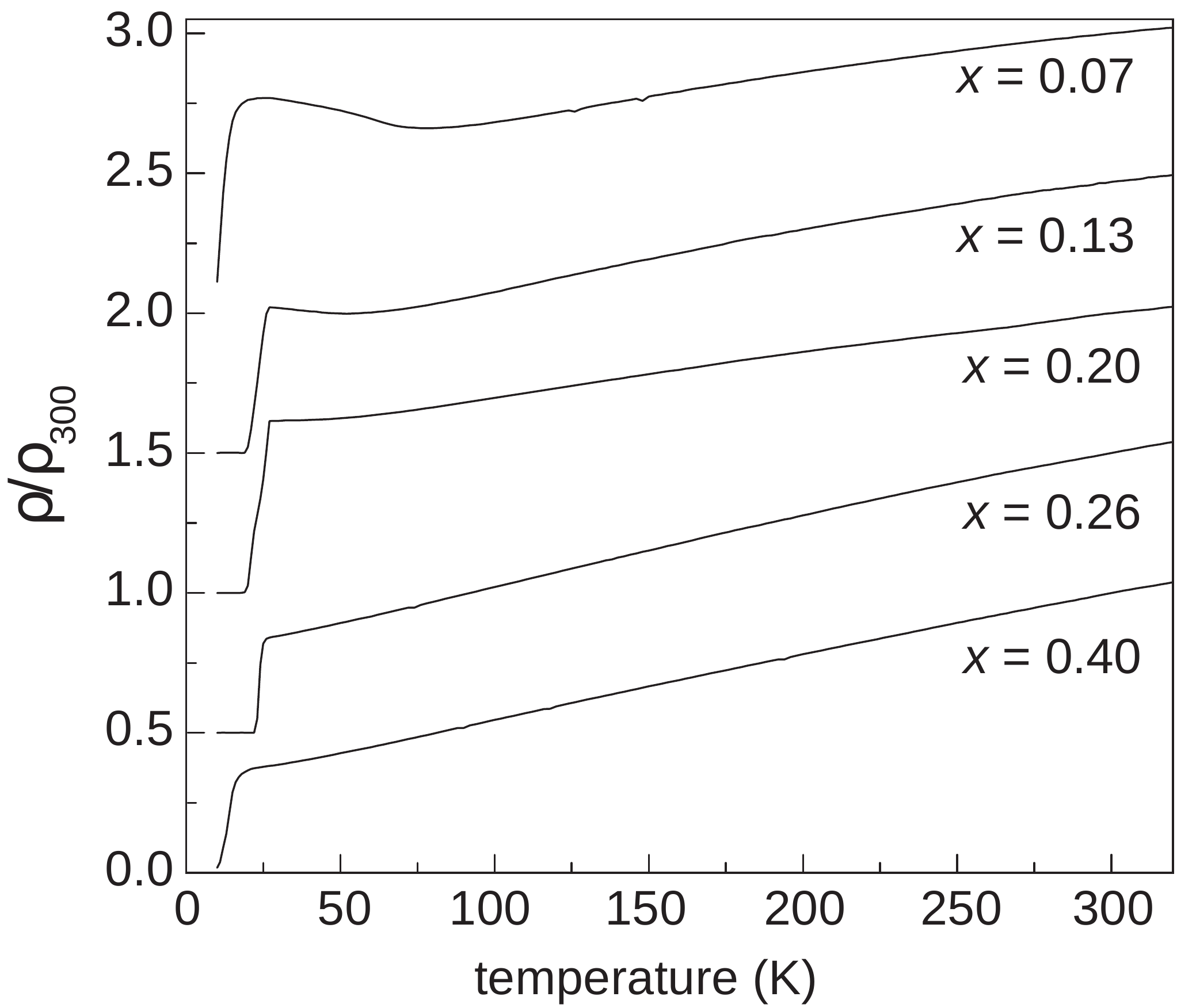}
\caption{{ \BadP}: Normalized electrical resistances $\rho$/$\rho_{300}$ for x=0.07-0.4.}
\label{fig:leit3}
}
\end{figure}

\begin{figure}[h!]
\center{
\includegraphics[width=0.6\textwidth]{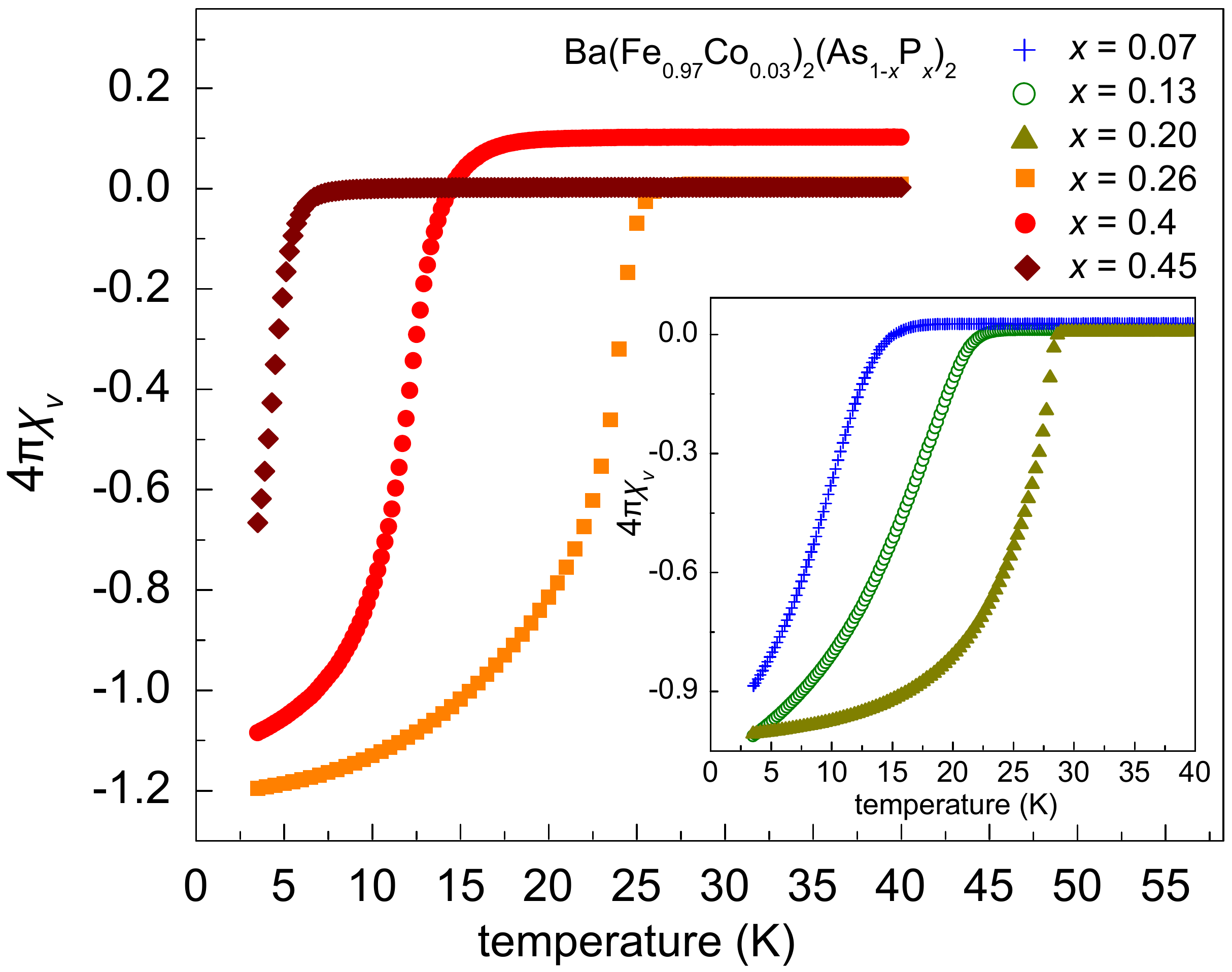}
\caption{{ \BadP}: ac-susceptibility $\chi$'$_v$ for x=0.07 -0.20 (blue-green, insert) and x=0.26-0.45 (orange-red, main plot).}
\label{fig:sus3}
}
\end{figure}

Although Co doping causes no significant structural changes, distinct effects on the physical properties occur. Figure \ref{fig:leit3} depicts resistivity versus temperature plots for the series {\BaCoP} with $y$ = 0.03 cobalt doping and $x$ = 0.07-0.40 while figure \ref{fig:sus3} presents the results of ac-susceptibility measurements confirming bulk superconductivity. $y$ = 0.03 ist just outside the superconducting dome; A superconducting volume fraction of only 3~\% and a $T_c$(onset) = 3~K was observed  for $y$ = 0.0315 Co, while bulk superconductivity with $T_c$=~7~K was found for $y$ = 0.0366 Co \cite{Rotundu-2011}. Substituting only $x$ = 0.07 phosphorus for arsenic yields a superconductor with $T_c$=~14~K. EDX measurements reveal a Co concentration of $y$ = 0.0322 in this sample, but even if bulk superconductivity could already be possible at this doping level, the transition temperature observed here is much higher than for Co doping alone.

Further substitution of arsenic by phosphorus quickly rises $T_c$ to a maximum of 28~K at $x \approx$ 0.2 as shown in figure \ref{fig:sus3}~(insert), followed by decreasing critical temperatures of 25 K at $x \approx 0.26$, 15 K at $x \approx 0.4$ and finally 5 K at $x \approx 0.45$ (main plot).

\begin{figure}[h!]
\center{
\includegraphics[width=0.6\textwidth]{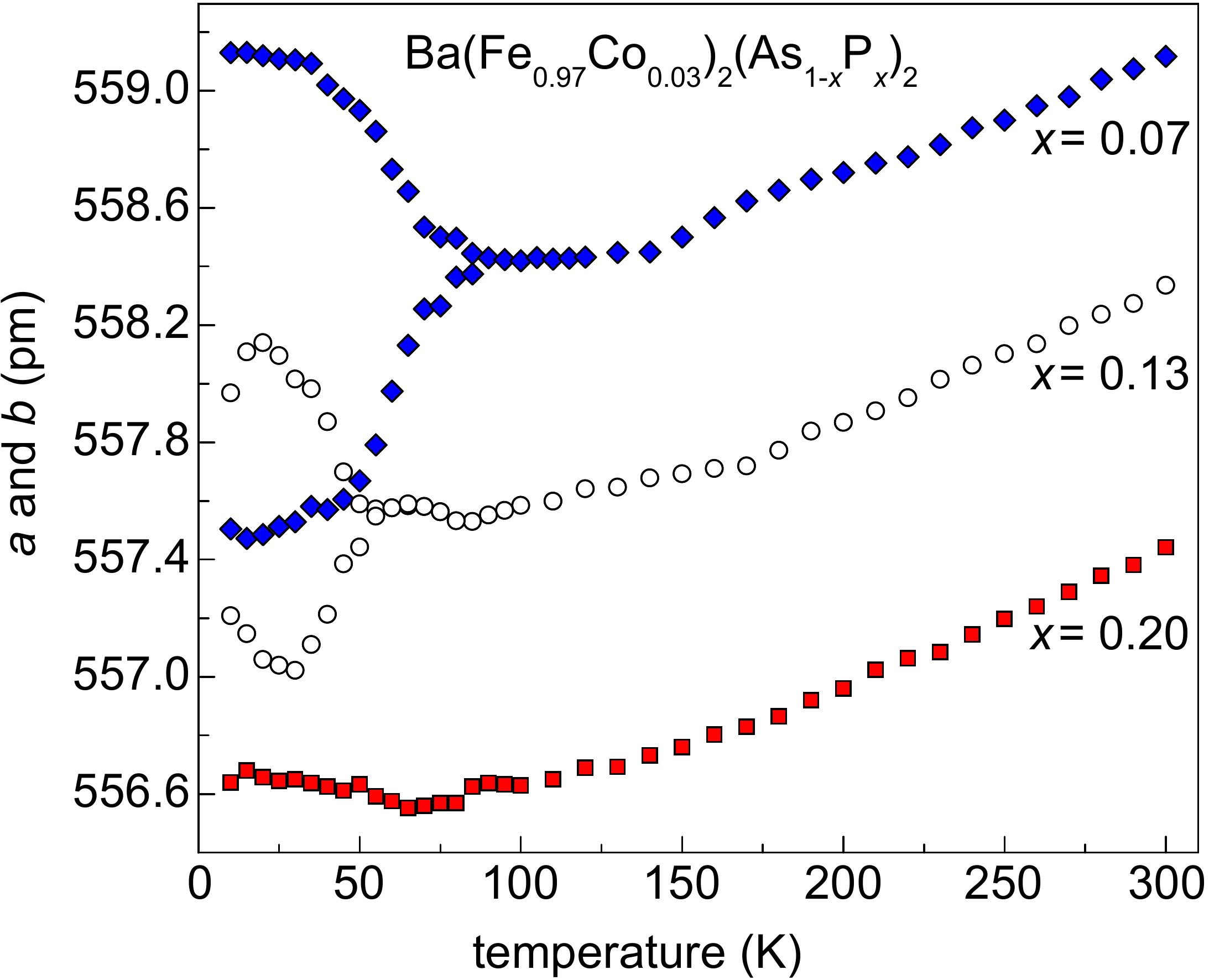}
\caption{{ \BadP}: temperature dependency of lattice parameters a and b (pm), for x=0.07 and 0.13 orthorombic distortion is observed.}
\label{fig:tieftemperatur}
}
\end{figure}

Samples with small phosphorus concentrations show resistivity anomalies around 90~K at $x$ = 0.07 and 70~K at $x$ = 0.13 (figure \ref{fig:leit3}).   Reports about underdoped {\BaCo} link an increase of the resistivity to the SDW-transition \cite{Ahilan-2008}. Although the effect is less pronounced   than reported for {\BaCo} single crystals, low temperature powder diffraction reveals a small splitting of reflections at $x$ = 0.07 and a broadening at $x$ = 0.13. Figure \ref{fig:tieftemperatur} shows Rietveld refinements of the low temperature powder data. The splitting of the lattice parameters appears close to 85~K at $x$ = 0.07 and near 60~K at $x$ = 0.13. Exact determinations of the transition temperatures are difficult due to peak broadening, possible phosphorus inhomogeneities and the smallness of the splitting. At 10~K the orthorhombic lattice parameters differ by only 1.6~pm at $x$ = 0.07 and by 0.7~pm at $x$ = 0.13 compared to 4~pm in BaFe$_2$As$_2$ \cite{Rotter-2008-1}.

\begin{figure}[h]
\center{
\includegraphics[width=0.6\textwidth]{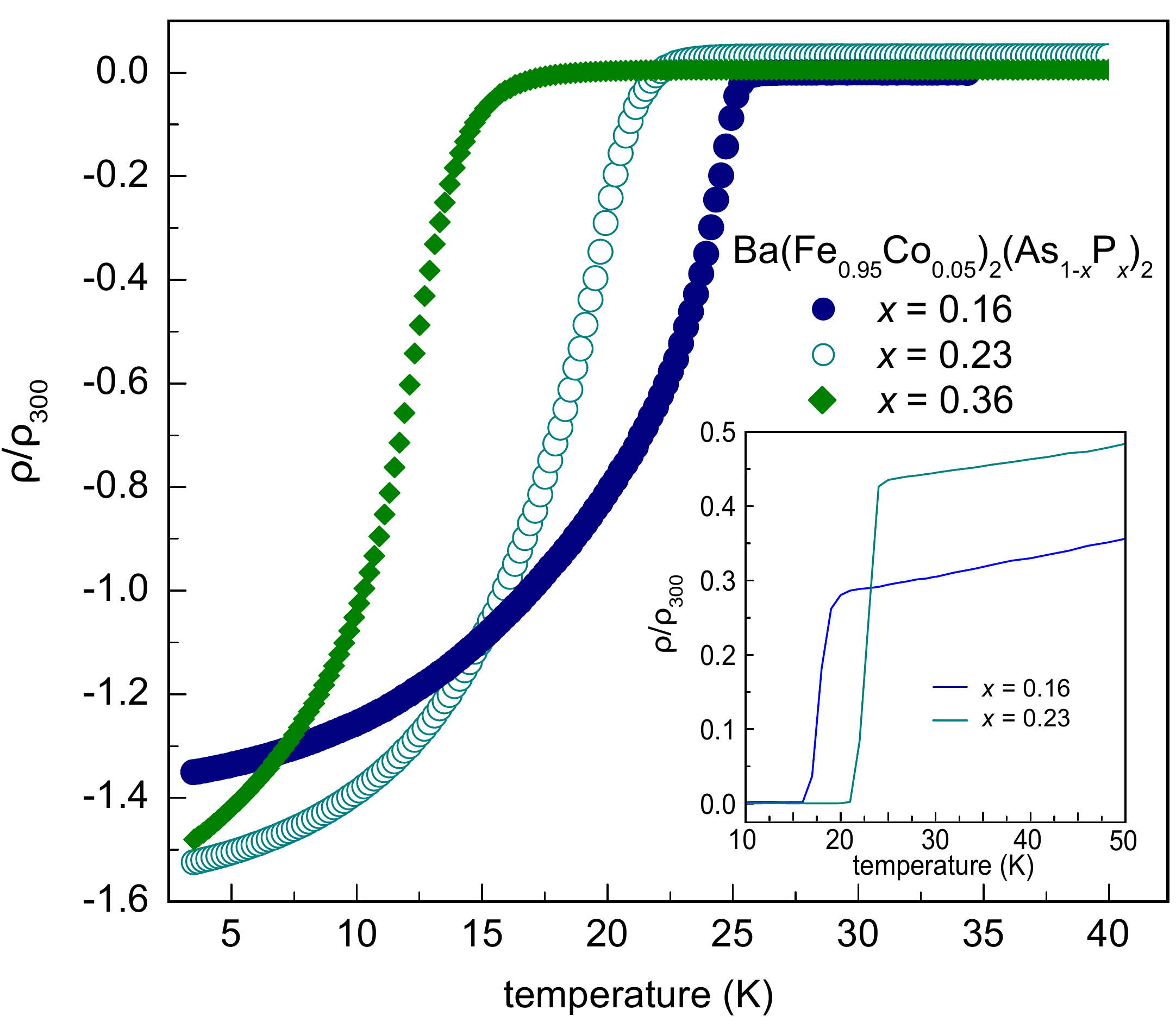}
\caption{ac-susceptibility of \BafP with x=0.16, 0.23 and 0.36; insert: relative resistance $\rho$/$\rho_{300}$.}
\label{fig:leitsus5}
}
\end{figure}

Next we studied \BaCoP~samples with the higher Co doping levels $y$ = 0.05 and $y$ = 0.07 that are already superconducting without phosphorus substitution. Figure \ref{fig:leitsus5} presents susceptibility and resistivity (insert) measurements for $x$ = 0.16-0.23. A transition temperature of 18 K was reported for Ba(Fe$_{0.95}$Co$_{0.05}$)$_{2}$As$_{2}$ \cite{Ahilan-2008} which increases to 24~K by phosphorus substitution at $x$ = 0.16 before dropping slightly to 22~K at $x$ = 0.23 and to 14.5~K at $x$ = 0.36. These effects are similar to those observed for the samples at $y$ = 0.03, though less pronounced.

\begin{figure}[h]
\center{
\includegraphics[width=0.6\textwidth]{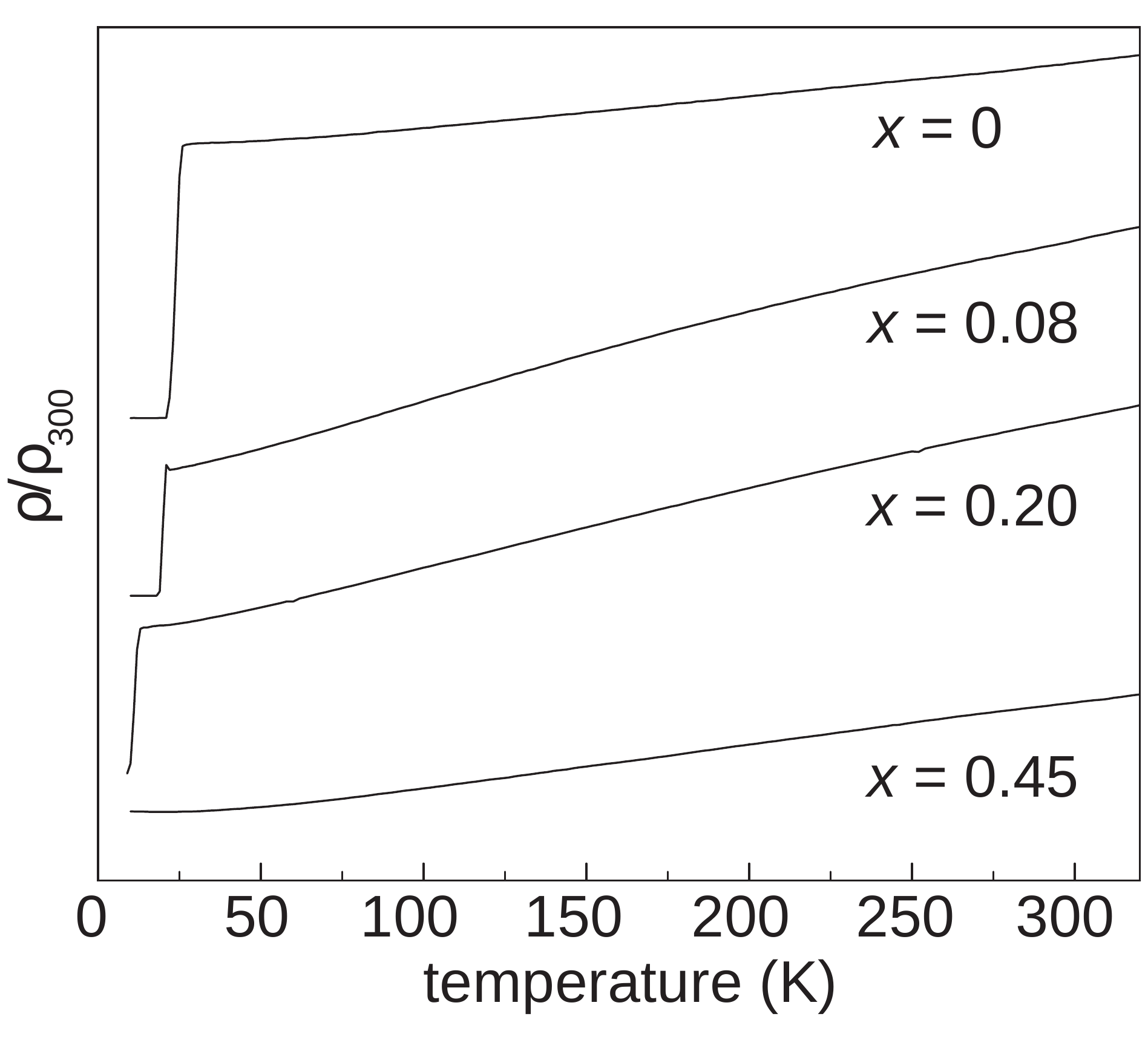}
\caption{{ \BasP}: Normalized electrical resistances $\rho$/$\rho_{300}$ for $x$ = 0-0.45.}
\label{fig:leit7}
}
\end{figure}

As soon as the Co doping level is increased to $x$ = 0.07, the picture changes. From the normalized electrical resistances depicted in figure \ref{fig:leit7} it becomes immediately clear that phosphorus-substitution in  Ba(Fe$_{0.93}$Co$_{0.07}$)$_{2}$As$_{2}$ always diminishes $T_c$ until superconductivity is suppressed. This is supported by ac-susceptibility measurements that show bulk superconductivity for $x$ = 0.08 and $x$ = 0.20 ($T_c$ = 21 and 13~K) while for $x$ = 0.32 only the onset of diamagnetism with a small superconducting volume fraction is detected.

\begin{figure}[h!]
\center{
\includegraphics[width=0.6\textwidth]{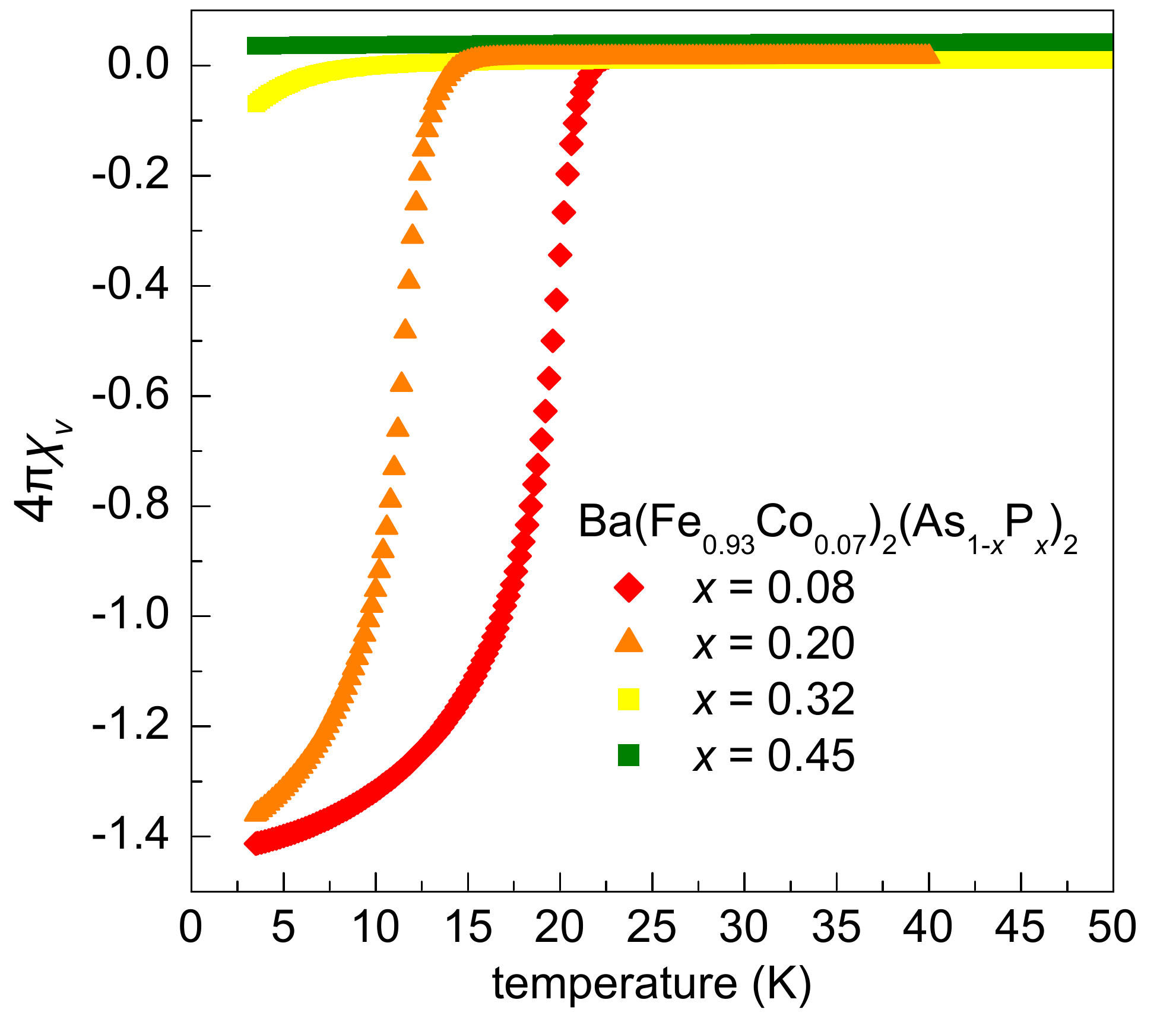}
\caption{ac-susceptibility of {\BasP} with x=0.08-0.45}
\label{fig:sus7}
}
\end{figure}

\begin{figure}[h]
\center{
\includegraphics[width=0.75\textwidth]{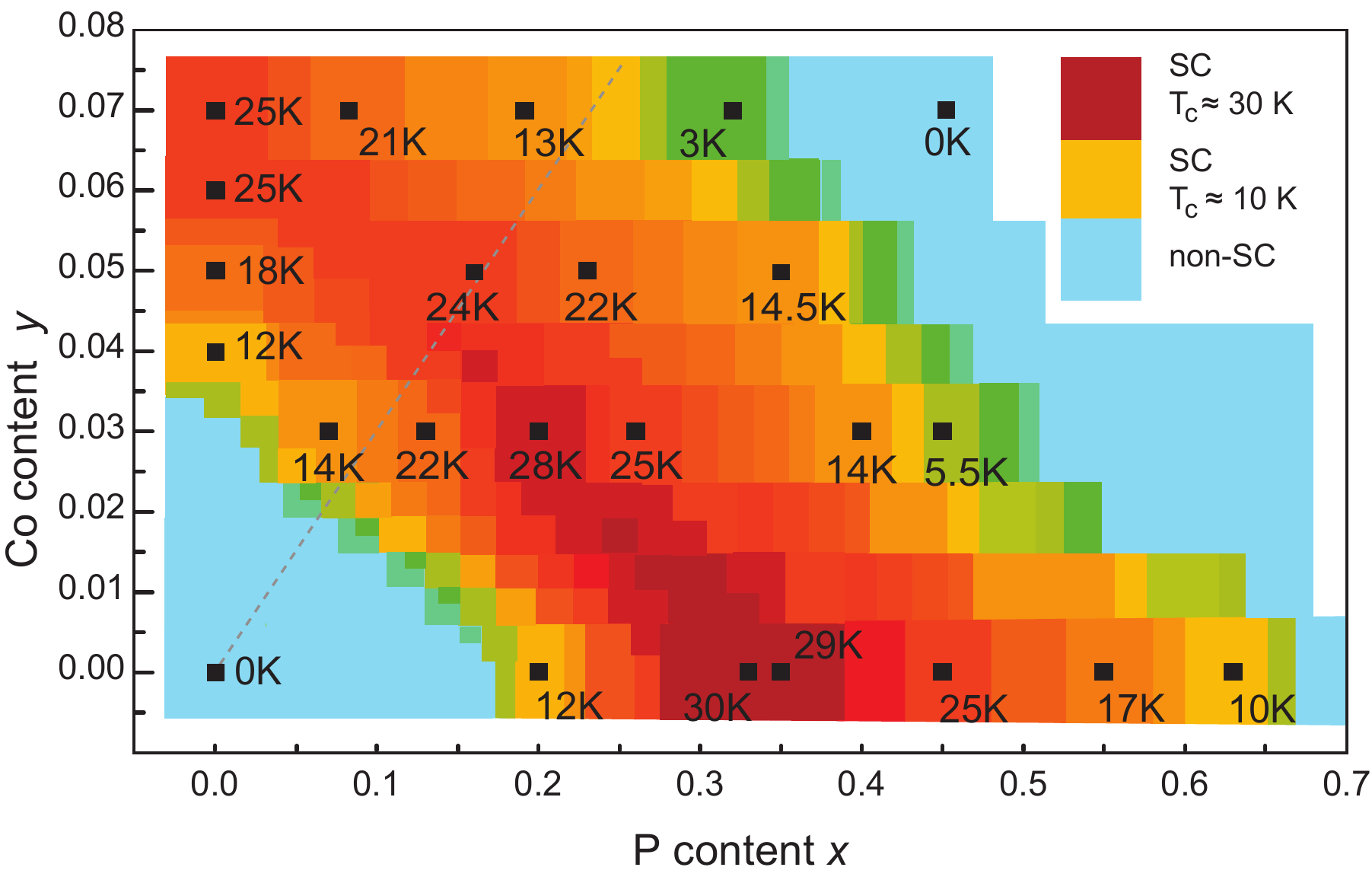}
\caption{Phase diagram of {\BaCoP}, data points in black, $T_c$ is colour-coded: $T_c \approx$ 30~K (dark red), $T_c \approx$ 10~K (yellow), non superconducting (blue); data for $x$ = 0 and $y$ = 0 are from \cite{Ni-2008} and \cite{Jiang-2009}}
\label{fig:phase3D}
}
\end{figure}

To summarize the results presented so far, we find an increase followed by a decrease of $T_c$ in cobalt-underdoped samples ($y$= 0.03 and 0.05, more pronounced in $y = 0.03$), and a decline of $T_c$ for the optimally doped ($y$ = 0.07) compounds upon phosphorus substitution. All data points are collected in a phase diagram (figure \ref{fig:phase3D}), where $T_c$ is colour-coded: dark red means $T_c \approx $ 30~K, yellow $T_c \approx$ 10~K and non superconducting areas are depicted in light blue. The data for {\BaCo} and {\BaP} were taken from Ref. \cite{Ni-2008} and Ref. \cite{Jiang-2009}. We see that the superconducting dome shifts to lower phosphorus contents if Co doping increases. Co and P doping seem to work together at suppressing the SDW transition, inducing superconductivity and increasing $T_c$  in the  underdoped areas, respectively, where neither one dopant alone could explain the observed $T_c$'s. However, Co and P doping together can not surpass the maximum $T_c$ of 30~K found in {\BaP} at $x$~=~0.33. Our results suggest that the maximal transition temperature of each Co doping level is shifted along a line connecting $x$ = 0.33 (P) and $y$ = 0.07 (Co), gradually decreasing from 30~K to 25~K.

Recent ARPES results suggested that phosphorus substitution may act like hole doping in {\BaP} with 0.3 holes per iron when one arsenic is replaced by phosphorus at $x$ = 0.5 \cite{Ye-2011}. In this case the same number of holes and electrons should be present along the dashed line in figure \ref{fig:phase3D}. However, our recent study of Ba$_{1-x}$K$_x$(Fe$_{1-y}$Co$_{y}$)$_{2}$As$_{2}$ shows a compensation of hole and electron doping and the recovery of a parent-like, non-superconducting phase for low cobalt doping \cite{Zinth-2011}. As can be seen from figure \ref{fig:phase3D}, no such effect is found for {\BaCoP}.

\begin{figure}[h]
\center{
\includegraphics[width=0.6\textwidth]{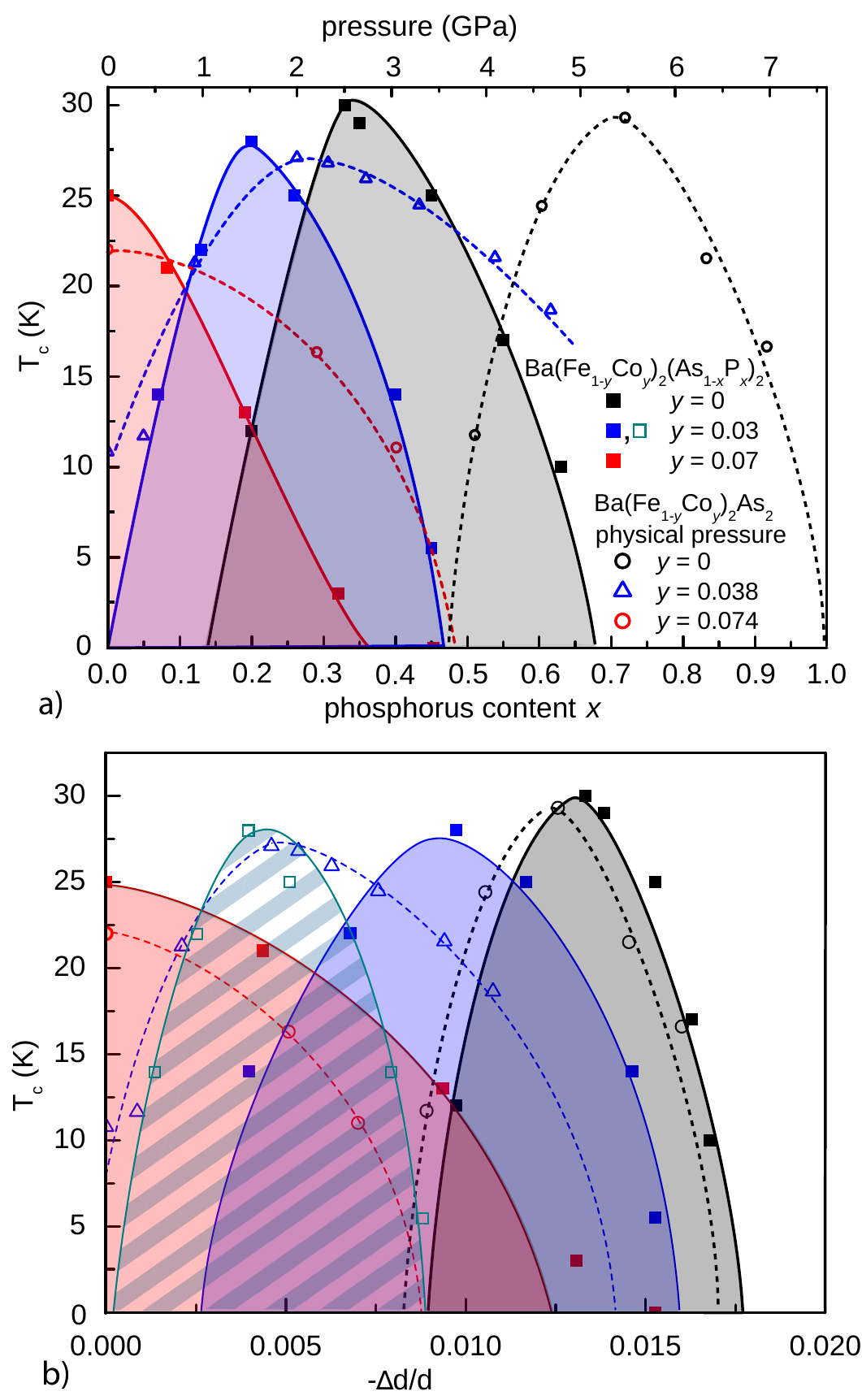}
\caption{a) Comparison of the effects of physical pressure on Ba(Fe$_{1-y}$Co$_{y}$)$_{2}$As$_{2}$ and chemical pressure in {\BaCoP};  continous lines, filled symbols and areas: {\BaCoP} with $y$ = 0 (black), $y$~=~0.03 (blue) and $y$~=~0.07 (red). Open symbols and dashed lines: {\BaCo}, physical pressure: $y$~=~0 (black) \cite{Colombier-2009}, $y$~=~0.038 (blue) and $y$~=~0.074 (red) \cite{Colombier-2010}. b) Plot of the same data relative to the change of the Fe-As distances taken from Figure \ref{fig:FeAs}. The green banded dome results if the linear decrease of the bond lengths from Fig.~\ref{fig:FeAs} is used.}
\label{fig:pressure}
}
\end{figure}

The structure of {\BaCoP} is dominated by the chemical pressure generated by phosphorus doping, therefore the comparison to physical pressure suggests itself. Several studies with {\BaCo} report the increase of $T_c$ under pressure in underdoped samples and mostly decrease for optimally doped ones \cite{Ahilan-2008, Colombier-2010, Drotziger-2010, Nakashima-2010}. Especially uni\-axial pressure seems to have a big influence, for example in Ba(Fe$_{0.92}$Co$_{0.08}$)$_{2}$As$_{2}$, where $T_c$ rises slightly with pressure under hydrostatic conditions, while it is quickly diminished if pressure is applied along the $c$-axis \cite{Nakashima-2010}. However, as phosphorus substitution intrinsically provides an uniaxial component \cite{Klintberg-2010}, our data should agree better with studies preformed under less hydrostatic pressure conditions. For comparison, we have related the phosphorus content to physical pressure via the cell volume, which gives 1~GPa ${\mathrel{\widehat{=}}}$ 13.12~\% phosphorus (data for {\BaP} from \cite{Martin-Thes}, and for pressurized  {\BaFA} from \cite{Jorgensen-2010}, ratio of changes for $c$ and $a$ similar for both data sets). Plots of $T_c$ versus phosphorus content (filled symbols) or physical pressure (open symbols and dashed lines) for different Co doping are displayed in figure \ref{fig:pressure}a ($y$ = 0 (black), $y \approx$ 0.03 (blue) and $y \approx$ 0.07 (red)).

The trend for each Co doping level is the same both for chemical and physical pressure. However, physical pressure produces broader superconducting domes which are shifted to higher pressures. It appears as if chemical pressure is more efficient in pushing the system to the maximum critical temperature  $T_{c,max}$, emphasized by the large pressure difference  of $\approx$ 3 GPa between {\BaFA} under physical pressure and {\BaP}.

\begin{figure}[h]
\center{
\includegraphics[width=0.7\textwidth]{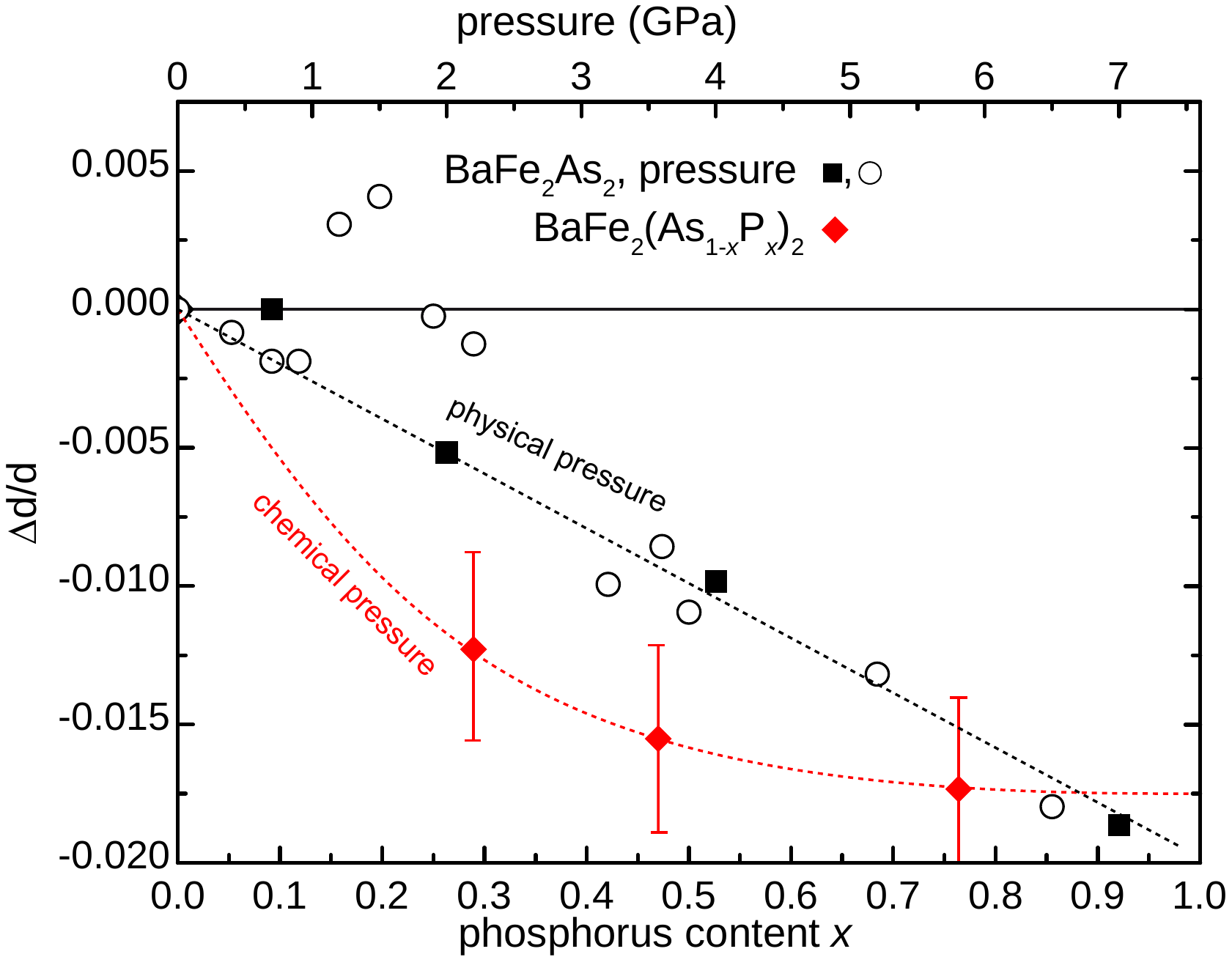}
\caption{changes in the normalized Fe-As bond length in {\BaP} \cite{Rotter-2010} (red) and {\BaFA} under pressure (black) \cite{Jorgensen-2010,Mittal-2011}, lines are guide to the eye, chemical pressure related to physical pressure via cell volume.}
\label{fig:FeAs}
}
\end{figure}

Considering details of the crystal structure helps to understand these discrepancies. While the lattice parameters $a$ and $c$ vary similar under chemical and physical pressure, significant differences are found in the reduction of the Fe-As bond lengths. Figure \ref{fig:FeAs} depicts the changes in the norma\-lized Fe-As distances in {\BaP} (red) \cite{Rotter-2010} and {\BaFA} under pressure (black) \cite{Jorgensen-2010,Mittal-2011}. While the Fe-As bonds in {\BaFA} decrease linearly with increasing pressure, the Fe-As bonds in {\BaP} contract rather strongly at low phosphorus doping levels and then converge \cite{Rotter-2010}.

The shortening of the Fe-As bond length is known to be connected with the suppression of the SDW-state in {\BaP}, thus chemical and physical pressure may result in similar states for similar Fe-As distances. For example, in {\BaFA} the Fe-As distance is reduced by $\approx$ 1.35~\% at $\approx$ 5.5~GPa, where $T_{c,max}$ is observed. An equal reduction of the bond length is found for $x\approx$~0.3 in {\BaP}, again close to the maxi\-mum of the superconducting dome. Figure \ref{fig:pressure}b shows a plot of $T_c$ versus the changes in the normalized Fe-As distance for the data depicted in \ref{fig:pressure}a, where a correlation between phosphorus content and physical pressure with the Fe-As distances was done by using the dashed lines of figure \ref{fig:FeAs}. This shows very good agreement between {\BaFA} and {\BaP}, indicating that, regardless of physical or chemical pressure, the Fe-As bond length is the crucial factor that controls $T_c$ if no Co doping is present.

For the Co underdoped sample ($y$ = 0.03) the maximal $T_c$ values for chemical and physical pressure coincide at the same value $-\Delta d/d \approx 0.005$ (green banded dome in Figure \ref{fig:pressure}b) only if we anticipate that the bond shortening with phosphorus substitution is linear in both cases. This assumes that the steeper decrease of $-\Delta d/d$ at lower phosphorus concentrations (figure \ref{fig:FeAs}) is due to the suppression of the magnetism, which may already be weakened in the case of Co doping. If we alternatively anticipate that the bond length curve for chemical pressure is equally valid for the Co doped sample, the maximum $T_c$
for chemical pressure is shifted to $-\Delta d/d \approx 0.01$ (blue filled dome in Figure \ref{fig:pressure}b), but the shape of the dome is now more similar to that of physical pressure.
A shift of the $T_{c,max}$ value for chemical pressure (filled blue) to higher $-\Delta d/d$ relative to $T_{c,max}$ for physical pressure (dashed blue) is expected due to the slightly lower Co concentrations (chemical pressure: $y$ = 0.03, physical pressure $y$ = 0.038). However, the observed shift is bigger than estimated by assuming a linear dependency $-\Delta d/d(T_{c,max})$, but a more concrete analysis is still not possible due to the lack of sufficiently precise structural data.

\section{Conclusion}

We have studied the effects of simultaneous doping with cobalt and phosphorus in {\BaCoP} and presented a comprehensive phase diagram. Superconductivity can be induced by a combination of both dopants in such areas where the content of only one would not suffice, but a combination of both can also drive the system from the optimally to an overdoped state. The maximum $T_c$ is not in the Co plus P doped area of the phase diagram, but occurs in {\BaP} at $x$ = 0.33. Our phase diagram gives no evidence of holes in {\BaP} as suggested by recent angle resoled photoemission experiments. Chemical and physical pressure act similarly in {\BaCo}, but our data strongly suggest that the most important parameter is the length of the Fe-As bond and not the unit cell volume. This shows that differences between chemical and physical pressure that manifest oneself as the non-linear reduction of the Fe-As distance in {\BaP} are strongly linked to the superconducting properties also in the Co doped materials.

\section{Acknowledgments}
This work was financially supported by the German Research Foundation (DFG) within the priority program SPP1458 (Grant JO257/6-1).

\section*{References}

\bibliographystyle{iopart-num}

\bibliography{BaCoP}

\end{document}